\documentclass[final,3p,times]{elsarticle}

\usepackage{amssymb}
\usepackage{bm}
\usepackage[hidelinks]{hyperref}
\usepackage[font=footnotesize]{caption}
\usepackage{siunitx}
\usepackage{float}
\usepackage{booktabs}
\usepackage{graphicx}
\usepackage{soul}
\graphicspath{{Figures/}}

\journal{*Corresponding author: \href{mailto:i.zivkovic@tue.nl}{i.zivkovic@tue.nl} }

\begin{document}

\begin{frontmatter}

\title{Twisted Pair Transmission Line Coil - A Flexible, Self-Decoupled and Robust Element for 7T MRI}

\author[inst1]{Jules Vliem}
\author[inst2,inst3]{Ying Xiao}
\author[inst2,inst3]{Daniel Wenz}
\author[inst2,inst3]{Lijing Xin}
\author[inst4]{Wouter Teeuwisse}
\author[inst4]{Thomas Ruytenberg}
\author[inst4]{Andrew Webb}
\author[inst1]{Irena Zivkovic*}

\affiliation[inst1]{organization={Department of Electrical Engineering, Eindhoven University of Technology},
            city={Eindhoven}, 
            country={The Netherlands}}

\affiliation[inst2]{organization={CIBM Center for Biomedical Imaging},
            city={Lausanne},
            country={Switzerland}}

\affiliation[inst3]{organization={École Polytechnique Fédérale de Lausanne (EPFL), Animal Imaging and Technology},
            city={Lausanne},
            country={Switzerland}}
            
\affiliation[inst4]{organization={C.J. Gorter MRI Centre, Department of Radiology, Leiden University Medical Center},
            city={Leiden},
            country={The Netherlands}}

\begin{abstract}
\textbf{Objective:} This study evaluates the performance of a twisted pair transmission line coil as a transceive element for 7T MRI in terms of physical flexibility, robustness to shape deformations, and interelement decoupling.\\
\textbf{Methods:} Each coil element was created by shaping a twisted pair of wires into a circle. One wire was interrupted at the top, while the other was interrupted at the bottom, and connected to the matching circuit. Electromagnetic simulations were conducted to determine the optimal number of twists per length (in terms of B$_1^+$ field efficiency, SAR efficiency, sensitivity to elongation and interelement decoupling properties) and for investigating the fundamental operational principle of the coil through fields streamline visualization. A comparison between the twisted pair coil and a conventional loop coil in terms of B$_1^+$ fields, maxSAR\textsubscript{10g}, and stability of $S_{11}$ when the coil was deformed, was performed. Experimentally measured interelement coupling between individual elements of multichannel arrays was also investigated.\\
\textbf{Results:} Increasing the number of twists per length resulted in a more physically robust coil. Poynting vector streamline visualization showed that the twisted pair coil concentrated most of the energy in the near field. The twisted pair coil exhibited comparable B$_1^+$ fields and improved maxSAR\textsubscript{10g} to the conventional coil but demonstrated exceptional stability with respect to coil deformation and a strong self-decoupling nature when placed in an array configuration. \\
\textbf{Discussion:} The findings highlight the robustness of the twisted pair coil, showcasing its stability under shape variations. This coil holds great potential as a flexible RF coil for various imaging applications using multiple-element arrays, benefiting from its inherent decoupling.

\end{abstract}



\begin{keyword}
Twisted Pair \sep Flexible Coils \sep 7T Arrays \sep Ultra-High Field MRI

\end{keyword}

\end{frontmatter}


\section{Introduction}
\label{sec:sample1}

Multi-channel RF coil arrays are ubiquitous in clinical and ultra-high field MRI and can be used for both RF transmission and signal reception~\cite{Oezerdem2016, Elabyad2021, Clement2019, Williams2021, Vaughan2004, Steensma2019}. In transmit mode, they can be advantageous for RF shimming~\cite{Curtis2012}, whereas, in a receive mode, they can provide a high Signal-to-Noise Ratio (SNR) and enable parallel imaging strategies~\cite{Adriany2005, Pruessmann1999, Griswold2002}. C]. Common building elements of such arrays are usually constructed using resonant copper loop elements, segmented with capacitors, which are fixed in space. In this approach, the array is designed such that it can fit a large population of patients. This is often associated with a reduced filling factor of an array leading to sub-optimal coil performance. Highly flexible RF coil arrays can be considered a very promising solution to address this limitation, and therefore there has been an ongoing interest in developing novel strategies for such flexible arrays~\cite{Ruytenberg2020, VanLeeuwen2022, Nohava2021, Port2020, Hosseinnezhadian2018}. The appeal of flexible coils lies in their ability to conform to diverse anatomies, maintaining a consistent distance from the subject and thereby improving electromagnetic (EM) performance through enhanced coupling to the sample~\cite{Nohava2021}. Additionally, flexible RF coil arrays improve patient comfort due to their lightweight and form-fitting design, making them well-aligned with the goal of creating patient-centred solutions.

Interelement coupling is considered a major challenge in any type of multichannel array; insufficient decoupling between the elements can lead to decreased performances of the array by transferring energy to adjacent elements instead of the subject. For accelerated imaging, it is essential to have as low coupling as possible. Ideally, each element should have a distinct sensitivity map for optimal acceleration ~\cite{Pruessmann1999, DeZwart2002}. When coils couple inductively, they become sensitive to the same regions of the sample, yielding intertwined sensitivity maps and diminished accelerated imaging efficiency~\cite{Weiger2001a, Ohliger2006a}.

Various methods are used for interelement decoupling for loop coils such as partial overlapping~\cite{Roemer1990}, capacitive and inductive networks~\cite{Zhang2004, Bilgen2006, Lee2002a}, transformer decoupling~\cite{Shajan2014} and the use of strategically placed passive resonators~\cite{Avdievich2013, Yan2014}. For receive-only arrays, preamplifier decoupling is used~\cite{Roemer1990}. The objective of these decoupling techniques is to prevent unwanted "communication" between the coils, i.e. parasitic current induction which could lead to a decline in coil performance. Ideally, the coil elements utilised in an array configuration would inherently possess decoupling properties, negating the need for any of the aforementioned decoupling techniques. A proposed intrinsically decoupled element is the shielded-coaxial-cable (SCC) coil, exhibiting a highly decoupled nature per se - it is demonstrated that placing SCC elements in various array configurations keeps elements highly decoupled which allows high-performance imaging at 7T~\cite{Ruytenberg2020, Ruytenberg2020a}. Given that a coaxial cable is a transmission line, the question arises if other transmission line elements hold the potential for a flexible RF coil design.

In this work an alternative transmission line transceive element is proposed - the twisted pair transmission line coil. Prior research explored the use of the twisted pair coil solely as receive-only elements in both 3T~\cite{Maravilla2022, Maravilla2023} and 7T~\cite{Czerny2023} MRI systems. However, these previous studies did not extensively explore aspects such as decoupling properties, Specific Absorption Rate (SAR), transmission efficiency, and overall coil flexibility and robustness. This is mainly because the coil was used exclusively as a receive-only element. 

In this work, we compared the twisted pair coil with a conventional coil in terms of transmit field (B$_1^+$) efficiency, maximum SAR\textsubscript{10g}, robustness towards elongation and shape changes and interelement decoupling when placed in various array configurations. The intrinsic highly decoupled nature of the proposed coil has also been investigated and elucidated by considering the power flow via calculation of the Poynting vector.    


\section{Theory and Methods}
\label{sec:Methods}

\subsection{The Main Coil Concept}

\begin{figure}[H]
	\centering
	\captionsetup{justification=centering}
	\includegraphics[width=\columnwidth]{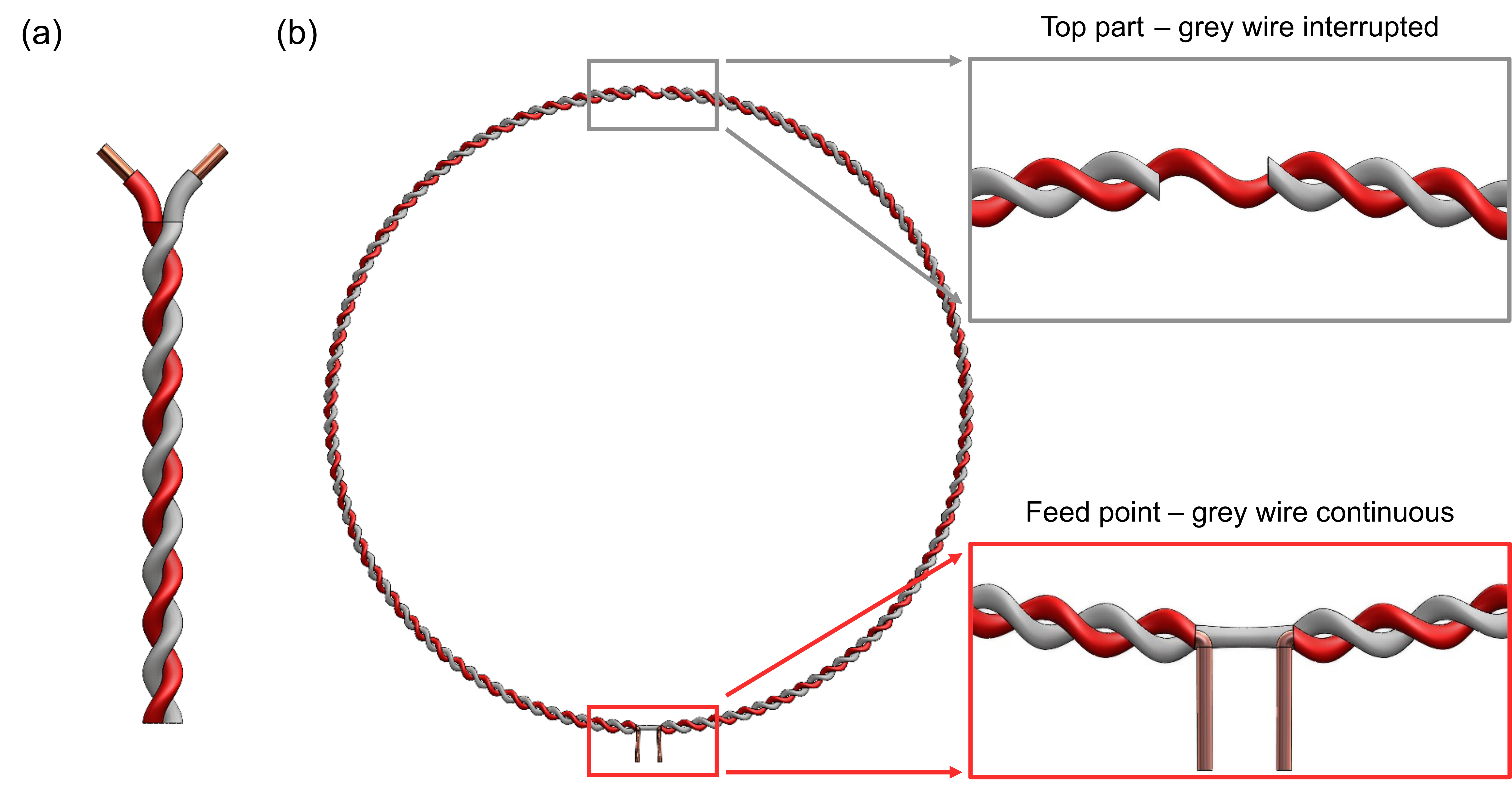}
	\caption{(a) Illustration of twisted pair cable. (b) Forming of the twisted pair coil. A gap is introduced in the grey \textit{shield wire} at the top to un-shield the magnetic field. In the red \textit{signal wire} a gap is introduced at the feeding point (bottom).}
	\label{fig:figure1}
\end{figure}

The twisted pair loop coil was created with two polytetrafluoroethylene (PTFE) insulated copper wires, twisted around one another, as can be seen in \autoref{fig:figure1}~(a). This created a flexible structure with a distributed capacitance. The twisted pair can be thought of as a parallel plate capacitor with the PTFE insulation of both wires as the dielectric. This distributed capacitance of the coil contains the electric field~\cite{Maravilla2023}. The twisted wire was shaped into a loop and gaps were introduced for both wires, as illustrated in \autoref{fig:figure1}~(b). The cut in the grey \textit{shield wire} - at the top, opposite of the feeding point, effectively un-shields the magnetic field, making the coil sensitive to electro-motive forces (EMF) and thus resonant. At the feeding point, a cut was made in the complementary red \textit{signal wire} for connection to the tuning and matching network, facilitating interfacing with the MRI scanner, and enabling the transmission and reception of signals.

When a current is excited on the red \textit{signal wire}, it flows in one direction, either clockwise or counter-clockwise. Take counter-clockwise as an example. Due to the tightly twisted configuration, a current is induced in the complementary grey \textit{shield wire}, flowing in the opposite direction according to Lenz's law, in this case clockwise. In the region in between the conductors of the wires, the magnetic fields generated by the currents point in the same direction, resulting from the opposite current flow in each wire. Outside the twisted wires, the magnetic fields tend to destructively interfere with each other since they flow in opposite directions.

Within the loop, a gap is introduced in the grey \textit{shield wire}, disrupting the continuous current path at that point. This gap is an open circuit, leading to a high voltage and a low current at the gap. A high potential difference is present between the two grey \textit{shield wire} ends, generating a strong electric field in the gap region. The strong electric field at the gap induces a substantial current in the red \textit{signal wire}, flowing in the same direction as the current in the grey \textit{shield wire}. This induced current contributes to the partial cancellation of the original current in the red \textit{signal wire} since it flows in the opposite direction (clockwise) to the original current (counterclockwise). The cancellation of the current in the red \textit{signal wire} results in the grey \textit{shield wire} being the dominant wire in determining the flow of the EM field. 
The grey \textit{shield wire} has no current close to the gap but here the red \textit{signal wire} takes over the current, resulting in a constant current flow in the clockwise direction for the loop. For our example the dominant current flow will thus be in the clockwise direction, creating a downward oriented magnetic field in the middle of the loop. Since an oscillating signal is used, the current flow and magnetic field will also switch, following the dominant current flow of the grey \textit{shield wire}. In \ref{appendix:APworkings} an illustration is shown to visualize the workings of the twisted pair coil.

\subsection{Simulation-Based Design of the Twisted Pair Coil}

The initial step in the design process involved determining the optimal twisting density per unit length for the wires. The twisted pair coil was modelled using SolidWorks (Dassault Systèmes SolidWorks Corporation, Waltham, USA) and imported in CST Studio Suite 2023 (Dassault Systèmes, Vélizy-Villacoublay, France) for electromagnetic simulations. The twisted pair was modelled by drawing two circles with 1 mm diameter, representing the copper conductor, spaced 2 mm apart. These circular profiles were swept along a predefined path, which could be circular or of arbitrary shape, to create the coil. The number of twists per unit length was specified in SolidWorks by the number of revelations of the profile along the path. All models have a total length of 31.4~cm, corresponding to a loop with 10~cm diameter, and are shaped into circular, elongated, or arbitrary forms. Cuts of 2~mm were introduced, one at the top and one at the bottom, for the grey \textit{shield wire }and red \textit{signal wire}, respectively. A discrete port was connected between the gap at the feed point, i.e. the red \textit{signal wire}. Lumped elements were connected to the port in the RF circuit co-simulation to tune the coil to the resonance frequency (297.2 MHz) and match the coil's impedance to 50~$\Omega$. 

To accurately mesh the complex geometry of the twisted shape, tetrahedral meshing is preferred. Tetrahedral meshing enables rapid and accurate meshing of intricately curved geometries compared to hexahedral meshing, which would not accurately capture the twisted and curved structure in a reasonable number of cells~\cite{Shepherd2008}. Therefore, the frequency domain solver with tetrahedral meshing was employed, utilising adaptive mesh refinement with a minimum of 3 and a maximum of 8 passes. The simulations were conducted by placing a coil on a homogeneous cubic phantom ($\varepsilon_r = 67.9$, $\sigma = 0.48$ S/m, dimensions 40$\times$40$\times$40~cm) with the coils positioned 2~cm away from the phantom. The copper cores were modelled as annealed copper ($\sigma = 5.8\cdot10^7$ S/m), while the insulation of both wires was excluded due to the increased complexity of the model. Additional simulations were conducted with the insulation layer to confirm the negligible impact on the simulated results. All the simulation results were normalised to 1 W of accepted power.

To investigate the optimal number of twists per unit length, various performance metrics on the cubic phantom were examined, including the reflection coefficient of the coil element, transmit efficiency (B$_1^+$) measured at 5~cm depth in the phantom at coil center, maximum Specific Absorption Rate at 10 grams average (maxSAR\textsubscript{10g}), SAR efficiency at 5~cm depth (B$_1^+$ divided by the square root of the maxSAR\textsubscript{10g}), and the coupling coefficient ($S_{21}$) between two neighbouring elements. The mentioned parameters were analysed for coils designed with different numbers of twists per coil length, such as 0, 10, 24, 48 and 64 twists per 31.4 cm (the circumference of all the 10~cm diameter loop coils). See \autoref{fig:figure2}~(a) for an illustration of two wires with different twists per unit length.

To compare the performance of the twisted pair coil we conducted a comparison with a conventional loop coil featuring distributed capacitors, illustrated in \autoref{fig:figure3}~(b), on a cubic phantom. The copper loop coil was modelled as a 10~cm diameter loop and constructed using 1~mm diameter annealed copper wire.

To better understand the differences between the EM fields of the twisted pair and conventional coils, we visualised the flow of these fields using streamlines. Streamlines are a visualisation technique that can be used to represent the flow of a vector field. It has been shown previously that using streamlines can give a much simpler and more direct explanation compared to a circuit approach, whose results are very close to each other (order of 0.1\%) ~\cite{Shamonina}. The fields were exported from CST to MATLAB to generate the streamlines in the slice plane. The fields were exported from CST to MATLAB to generate the streamlines in the slice plane. This visualisation allowed us to determine the difference in electromagnetic flow between the conventional and twisted pair coil, as well as when two coils are placed close to each other.

To demonstrate the extraordinary flexibility of the twisted pair coil, we further expanded its design by modelling it as a spline shape. This approach allowed us to create diverse and intricate coil configurations by defining arbitrary splines in SolidWorks and sweeping the circular profiles along the spline path. The coil was tuned and matched at the beginning in a circular shape (diameter of 10 cm) and subsequently, the shape was deformed while keeping the same circumference and tuning and matching circuit elements. 

\subsection{Coil Construction}

The twisted pair coil was constructed by tightly twisting two 18-gauge wires (1~mm diameter) with PTFE insulation. To achieve a high number of twists per unit length, two large sections of wire were cut, and one end was secured in a bench vise. The other end of the wires was inserted into a drill, and by turning the drill, the wires were twisted while maintaining tension by gently pulling them. The maximum twists per unit length were reached when the wires became significantly taut. At this point, the twisted pair could be removed from the vise and drill. Twisted pair coils which we fabricated had 64 twists per coil length (31.4 cm).

Before connecting the loop to the circuitry, a cut was made in the middle of the grey \textit{shield wire}. The size of this cut affected the coil's tuning, where a larger cut slightly increases the resonance frequency. The grey \textit{shield wire} cuts were approximately 2~mm in our construction. The coil's physical realisation and its corresponding schematic are shown in \autoref{fig:figure3}~(a). The coil, with a diameter of 10~cm, was tuned and matched using one parallel and two series fixed capacitors (AVX800 E Series, Kyocera-AVX, Fountain Inn, USA). The red \textit{signal wire} was attached to both sides of the parallel capacitor, while the grey \textit{shield wires} were connected, either twisted or soldered on the same trace of the Printed Circuit Board (PCB). The PCB was created by milling signal traces on a copper-plated FR4 material ($\varepsilon_r=4.4$). The circuitry was fed by an RG58 coaxial cable (RG58 LSZH, Multicomp PRO, London, UK) connecting the core to the signal trace and the shield to the ground trace. 

To compare the performance of the twisted pair loop coil, a conventional copper loop was also fabricated, as depicted in \autoref{fig:figure3}~(b) along with its circuitry. The conventional coil was constructed using a copper wire with a wire diameter of 1~mm. Fixed and variable capacitors were distributed along the 100~mm diameter loop, one parallel and two series fixed capacitors were utilised for tuning to the resonance frequency and matching the coil's impedance to 50~$\Omega$.

\subsection{Array Construction}

To assess the performance of individual elements in an array configuration, three distinct array layouts were examined. The first configuration featured an eight-element (two-by-four) array. All elements were circular and exhibited partial overlap, i.e. neighbouring elements within the same row and between the two rows were overlapped. This arrangement aimed to accommodate all eight coils within the given phantom space. Notably, no precise overlap distances or metrics were employed, highlighting the inherent decoupling of the coils, even when arranged in a semi-random fashion. 

The second array configuration also adopted a two-by-four arrangement. Coils within each row were slightly elongated to fit the available space without overlapping neighbouring elements. The elements positioned between the two rows were rotated by 180 degrees relative to each other and demonstrated partial overlap, although no predefined overlap degree was stipulated. 

Lastly, a five-element array was constituted by positioning extremely elongated elements sequentially to fill the available phantom space.

\subsection{Bench and MRI Measurements}

\subsubsection{Bench Measurements}

The individual elements were assessed on a cubic phantom, see \autoref{fig:figure3}~(c), placed 2~cm away from the phantom using a foam spacer. $S$-parameters were measured with a Vector Network Analyser (VNA) (Keysight N9914A FieldFox VNA 6.5GHz) for individual elements and when placed in an array configuration.

To investigate the coupling behaviour ($S_{21}$) of the twisted pair coil, two elements were placed in close proximity to the phantom. The distance between the elements was systematically adjusted, starting from 20~mm apart and gradually reducing the separation until they overlapped by 90~mm (90\% of the coil's diameter). The same procedure was used for two conventional coils. Additionally, different geometry orientations were tested for the twisted pair coils while varying the degree of overlap.

To measure the quality factors of the coil the half-power bandwidth (HPBW) edges were found when looking at the input impedance ($Z_{in}$) of the fabricated coils. Dividing the central frequency by the HPBW gives the Q factor of the coil~\cite{Pozar2011}. To measure the unloaded Q, the coil was placed in the air and for the loaded Q it was placed on the phantom.

\subsubsection{MRI Measurements}

The rectangular body phantom ($\varepsilon_r$ = 50, $\sigma$ = 0.6 S/m, size = $40\times40\times30$~cm) experiments were performed on a 7T Philips Achieva scanner. The 2D B$_1^+$ maps were obtained using the dual refocusing echo acquisition mode (DREAM)~\cite{Nehrke2012} sequence (FOV = $350\times350\times54$~mm$^3$, slices = 5, flip angle = 10\textdegree, STEAM angle = 50\textdegree, TR/TE = 4.0/0.98 ms, scan time = 120~s). 

The spherical phantom ($\varepsilon_r = 75$, $\sigma = 1.09$ S/m, diameter = 17~cm) experiments were performed on a 7T MR human scanner (Magnetom, Siemens Healthineers, Erlangen, Germany). The three-dimensional B$_1^+$ maps were quantitatively measured with the SA2RAGE~\cite{Eggenschwiler2012} sequence (TR/TE = 2400/0.78 ms, TI1/TI2 = 45/1800 ms, $\alpha_1/\alpha_2$ = 4/10$^\circ$, FOV = 208×256 mm$^2$, slices = 64, resolution = 2.0×2.0×2.5 mm, BW = 1220Hz/px, scan time = 115 s). The measurement results were normalised to the input power.

Furthermore, additional simulations were performed on the spherical phantom to enable a comparison between measurement and simulation results. The coils conformed to the surface of the sphere, by projecting the shape on the phantom so that a constant distance of 5~mm to the phantom was maintained. Simulated results were normalised to 1~W of accepted power. 

\section{Results}\label{sec:Results}

\subsection{Individual Coil Design and Comparison}

The simulated metrics of twisted pair coils across varying twist densities are presented in \autoref{fig:figure2}. It is observed that increasing the number of twists per unit length decreases the B$_1^+$ field efficiency, maxSAR\textsubscript{10g}, impedance of the coil (the points move from the right to left on the Smith chart) and coupling coefficient. A similar trend is observed for the SAR efficiency, despite being less pronounced as it levels off at around 24 twists per unit length. The difference in simulated B$_1^+$ fields of coils with 24 twists and 64 twists per unit length was $\sim$15\%, whereas in the measurements the difference was approximately 25\%. Additional simulated robustness analysis results for different numbers of twists per unit length are provided in \ref{appendix:AProbustness}, and show the enhanced robustness of the 64 twists per unit length configuration. For this research, a coil with a low coupling coefficient and high robustness towards the shape deformation is preferred over a high field efficiency so the coil can be efficiently used in densely populated receive and transceive arrays. Therefore, from now on, simulated and manufactured coils are always with 64 twists per unit length.

The fabricated twisted pair, with 64 twists per unit length, and conventional coils with their corresponding circuitry are shown in \autoref{fig:figure3} together with the cubic phantom used for $S$-parameter measurements. Additional return loss measurements are presented in \ref{appendix:APmeasured}, showcasing the robustness of the twisted pair to elongations compared to the conventional coil. The measured unloaded ($Q_{\text{un}}$)/ loaded ($Q_{\text{lo}}$) quality factors of the twisted pair were 55.0/49.4 whereas those of the conventional coil were 127.4/21.1, corresponding to ratios of $Q_{\text{un}}/Q_{\text{lo}}$ of 1.1 and 6.0.

A comparison of simulated B$_1^+$ field efficiency, maxSAR\textsubscript{10g} and SAR efficiency of the twisted pair and conventional coil, for different degrees of elongation is shown in \autoref{fig:figure4}. Simulated B$_1^+$ fields of the twisted pair and conventional coils were very similar. Simulated maxSAR\textsubscript{10g} for a twisted pair coil in a circular shape was 1.1~W/kg and decreased to 1.0~W/kg in the two elongated shapes. The conventional coil had a maxSAR\textsubscript{10g} of 1.1 W/kg in the circular shape while maxSAR\textsubscript{10g} increased to 1.4~W/kg in the maximum elongated shape. When comparing the simulated and measured B$_1^+$ field patterns for different degrees of elongation and good match is observed. The main difference is that the simulations were scaled down with 25\%. The SAR efficiency at superficial depth was higher for the elongated coils, for both the twisted pair and conventional coil, where after approximately 2 cm all profiles became very similar.

\subsection{Coupling Results}

The measured coupling coefficients ($S_{21}$) between two coils for varying overlapping geometries can be seen in \autoref{fig:figure7}. Across all the examined scenarios, the decoupling between two twisted pair elements was below -13 dB for the twisted pair. When comparing the performance of the twisted pair and the conventional coil across different degrees of overlap, the measured $S_{21}$ for the twisted pair was consistently below or equal to -15 dB, whereas the conventional coil exhibited poorer performance with $S_{21}$ values below or equal to -4 dB. Notably, the highest interelement coupling for the twisted pair, reaching -13 dB, was observed when the coils were rotated 180 degrees relative to each other. In the remaining two examined scenarios, the coupling remained below or equal to -18 dB.


\subsection{Streamline Plots}


The electric ($\bm{E}$) and magnetic ($\bm{H}$) fields, together with the power flow (Poynting vector $\bm{S}$) for conventional and twisted pair coils without a phantom is depicted in \autoref{fig:NoPhantom} in three different cross-sections. The $\bm{E}$ field consistently maintains its pattern and remains zero in the YZ plane. In contrast, the $\bm{H}$ field exhibits a similar pattern but significantly greater magnitude in the inner part of the conventional coil compared to the twisted pair coil. $\bm{S}$ lines propagate radially within the plane of the conventional coil, while in the case of the twisted pair coil, power lines exit from the gap opposite the port.

In \autoref{fig:WithPhantom}, we observe $\bm{E}$ and $\bm{H}$ fields and power flow $S$ for conventional and twisted pair coils with a phantom, captured in three different cross-sections. The $\bm{E}$ field maintains its shape and magnitude, remaining zero in the YZ plane. Conversely, the $\bm{H}$ field undergoes changes, displaying a similar magnitude and shape for both coils and becoming non-zero in the XZ plane. Analogous to the power bubble described in~\cite{Shamonina}, we identify a magnetic field bubble here—a border surrounding the coil, where magnetic field lines approach from one side and recede from the other. Notably, the magnetic field bubble is considerably smaller in the twisted pair coil compared to the conventional coil, indicating that only a small portion of the magnetic field from the phantom couples with the twisted pair coil. In both cases, $\bm{S}$ streamlines are directed toward the phantom. Additionally, the conventional coil features $\bm{S}$ streamlines propagating on both sides within its plane. Introduction of the phantom leads to the emergence of a very strong $\bm{E}$ field and power lines between the twisted pair coil and the phantom, driven by significant capacitive coupling.

\autoref{fig:TwoWithPhantom} illustrates $\bm{E}$ and $\bm{H}$ fields, current density $\bm{J}$, and power flow $\bm{S}$ for pairs of conventional and twisted pair coils with a phantom, captured in three different cross-sections. In these experiments, both coil types were positioned at distances of 5~cm and 5~mm from each other, with one coil actively observing the coupling to the other (non-active) coil. When the coils were separated by 5~mm, an analysis of the current density at the phantom’s surface revealed that the second conventional coil (the passive one) exhibited significantly greater coupling than the twisted pair coil, inducing a stronger current in the phantom. Magnetic field lines encircle both the active and passive conventional coils at both separation distances due to stronger coupling between the coils.For the twisted pair coil, the second (passive) coil also creates a magnetic field bubble when spaced 5~mm apart but is considerably smaller than the bubble from the conventional coil. Unlike conventional coils, where power flows from the active coil towards the passive one, resulting in coil coupling, the power lines in twisted pair coils do not flow from the active towards the second passive coil.

\subsection{Array Results}

\autoref{fig:figure9} shows the experimentally measured interelement coupling of the three array configurations. In all evaluated cases the return loss $S_{ii}$ of individual elements as well as interelement decoupling, $S_{ij}$, had excellent performance ($\leq$ -10 dB). The worst coupling results were observed in configuration (b) – eight slightly elongated, partially overlapped elements – for the opposing neighbours.

\subsection{Spline Shape Evaluation}

In \autoref{fig:SplinesAll}, two spline twisted pair coil shapes are presented. The coils were tuned and matched in the circular shape and subsequently their shapes were deformed without retuning the coils. Both twisted pair spline coils stayed decoupled ($<$ -15dB) for any degree of overlap. The B$_1^+$ field of the spline-shaped coils followed the coil shape, as seen in the coronal view. The maxSAR\textsubscript{10g} of the first spline \textit{saddle-shaped} coil slightly increased to 1.2W/kg, compared to the circular shape, while the maximum SAR\textsubscript{10g} of the second spline \textit{triangular-shaped} coil was the same as the circular shape (1.1 W/kg).


\section{Discussion}\label{sec:discussion}

In this work, we presented a novel design for an extremely flexible and self-decoupled coil element for 7T MRI – the twisted pair transmission line coil. The proposed coil does not require distributed lumped elements and due to its intrinsic decoupling properties, it allows for straightforward construction of tight-fitting multi-channel arrays. In MRI experiments, the proposed coil was used as a transceive element and a good agreement between simulations and measurements was achieved. To the best of our knowledge, this is the first time that a twisted pair transmission line coil was examined as a transmit/transceive element at 7T MRI. 

The advantage of the twisted pair coil, in comparison to other highly flexible elements such as the liquid metal coil ~\cite{Port2020}], lies in its inherent self-decoupled nature. Unlike the liquid metal coil, which requires mutual overlap to geometrically decouple the array, the twisted pair coil does not require any additional decoupling methods. When comparing the twisted pair coil with the Shielded-Coaxial-Cable (SCC) coil ~\cite{Ruytenberg2020}, similar decoupling properties were observed. In \cite{Ruytenberg2020}, it is stated that the $Q_{\text{ul}}/Q_{\text{lo}}$ of the conventional loop element was 5.7, while for the SCC coil, it was 1.7. In our measurements, $Q_{\text{ul}}/Q_{\text{lo}}$ of the twisted pair coil was 1.1, even lower than that of the SCC coil and over five times lower than that of the conventional coil. If we consider that at around 300 MHz, resistive coupling of coil-to-coil ``through a sample`` is the dominant coupling mechanism ~\cite{Ruytenberg2020}], the results imply lower coupling to the sample of the twisted pair coil vs. SCC, which further implies lower inter‐element coupling of twisted pair coils. While Ruytenberg et al. did not report on the coupling behaviour of the SCC when the loops have different orientations with respect to each other, in our study the interelement decoupling properties were excellent, regardless of the loop’s relative orientation and amount of overlap. 

\begin{figure}[H]
	\centering
	\captionsetup{justification=centering}
	\includegraphics[width=\textwidth]{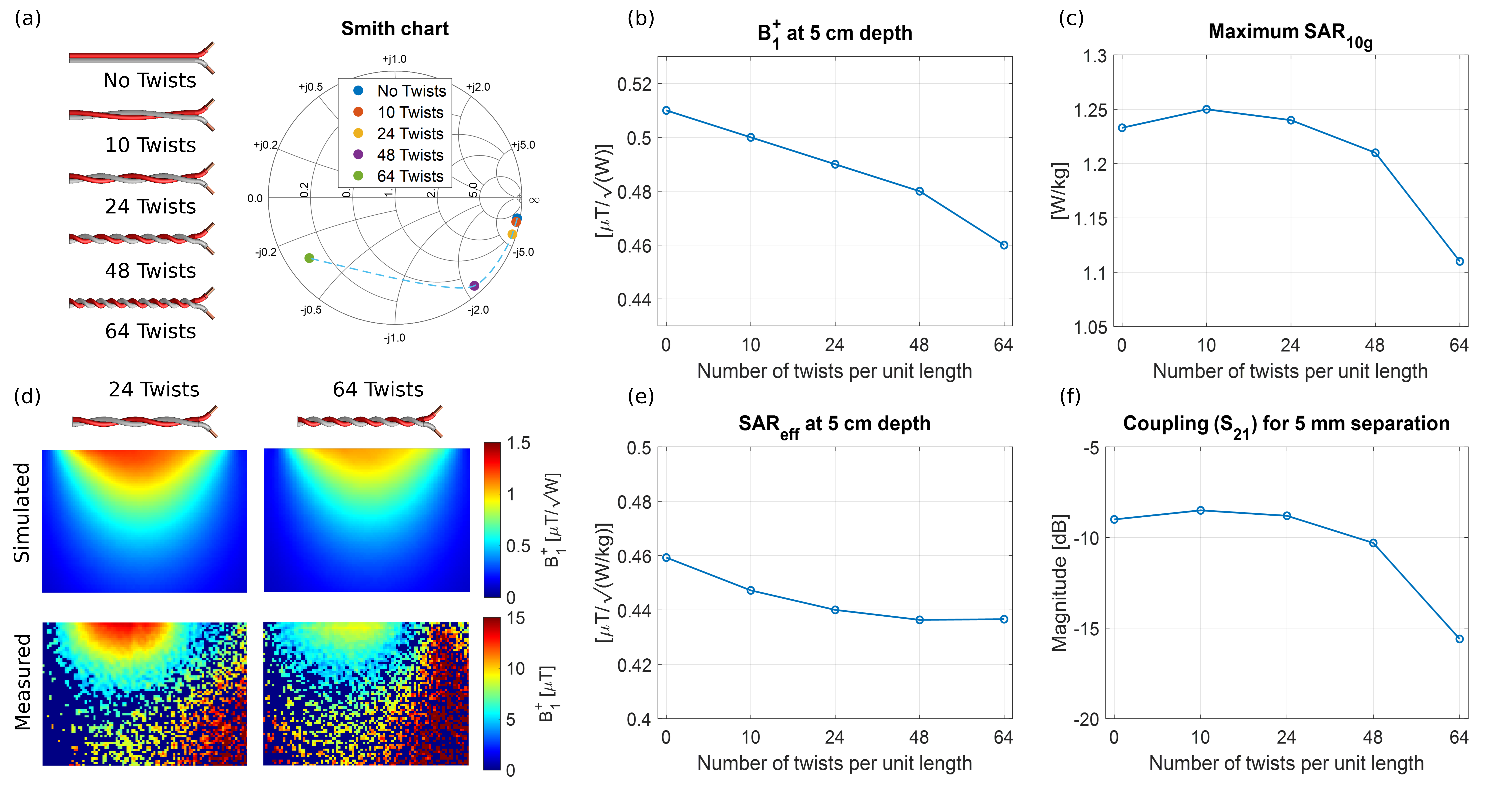}
	\caption{Simulated metrics of twisted pair coils across varying twisting densities. (a) Denoting number of twists per coil and its influence on the impedance of the coil, shown with reflection coefficients of the coil on the Smith chart. (b) Magnitude of B$_1^+$, in $\mu \text{T}/\sqrt{\text{W}}$, measured at 5~cm depth in the phantom at coil center. (c) Maximum SAR\textsubscript{10g} value, in W/kg, on the phantom. Both B$_1^+$ and maxSAR\textsubscript{10g} are normalised to 1~W accepted power. d) Simulated (top) and measured (bottom) B$_1^+$ field maps for 24 and 64 twists per length. (e) SAR efficiency: B$_1^+$ magnitude at 5cm depth divided by the square root of peak SAR, in $\mu \text{T}/\sqrt{\text{W/kg}}$. (f) Coupling coefficient ($S_{21}$), in dB, between two coils, spaced 5~mm apart.}
	\label{fig:figure2}
\end{figure}

\begin{figure}[H]
	\centering
	\captionsetup{justification=centering}
	\includegraphics[width=\textwidth]{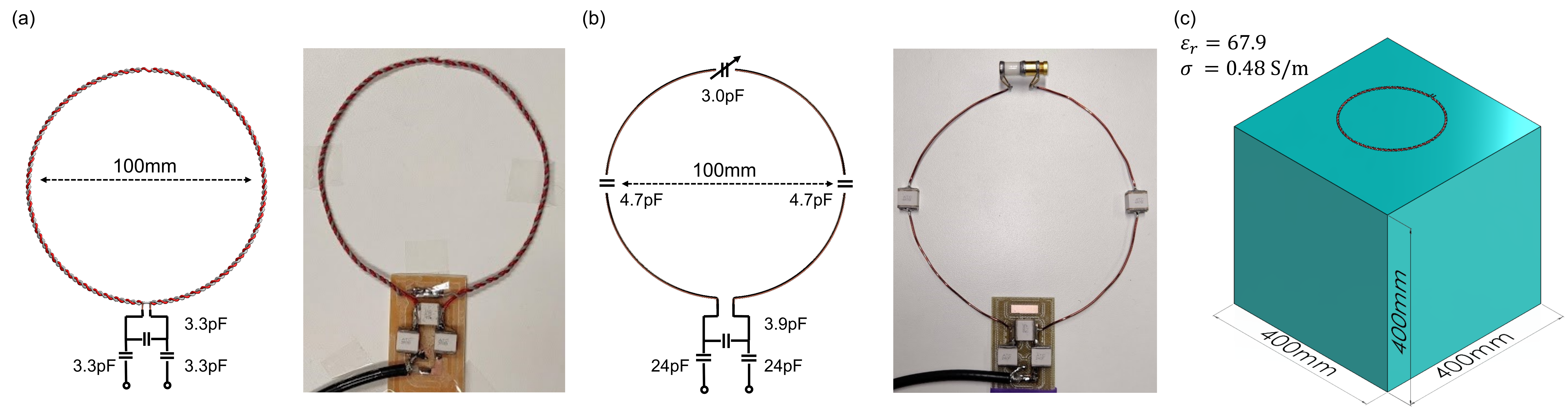}
	\caption{Schematics and photos of (a) twisted pair loop coil and (b) conventional loop coil. (c) Homogeneous cubic phantom used in simulations and bench measurements. The coil was placed 2~cm away from the phantom.}
	\label{fig:figure3}
\end{figure}

\begin{figure}[H]
	\centering
	\captionsetup{justification=centering}
	\includegraphics[width=\textwidth]{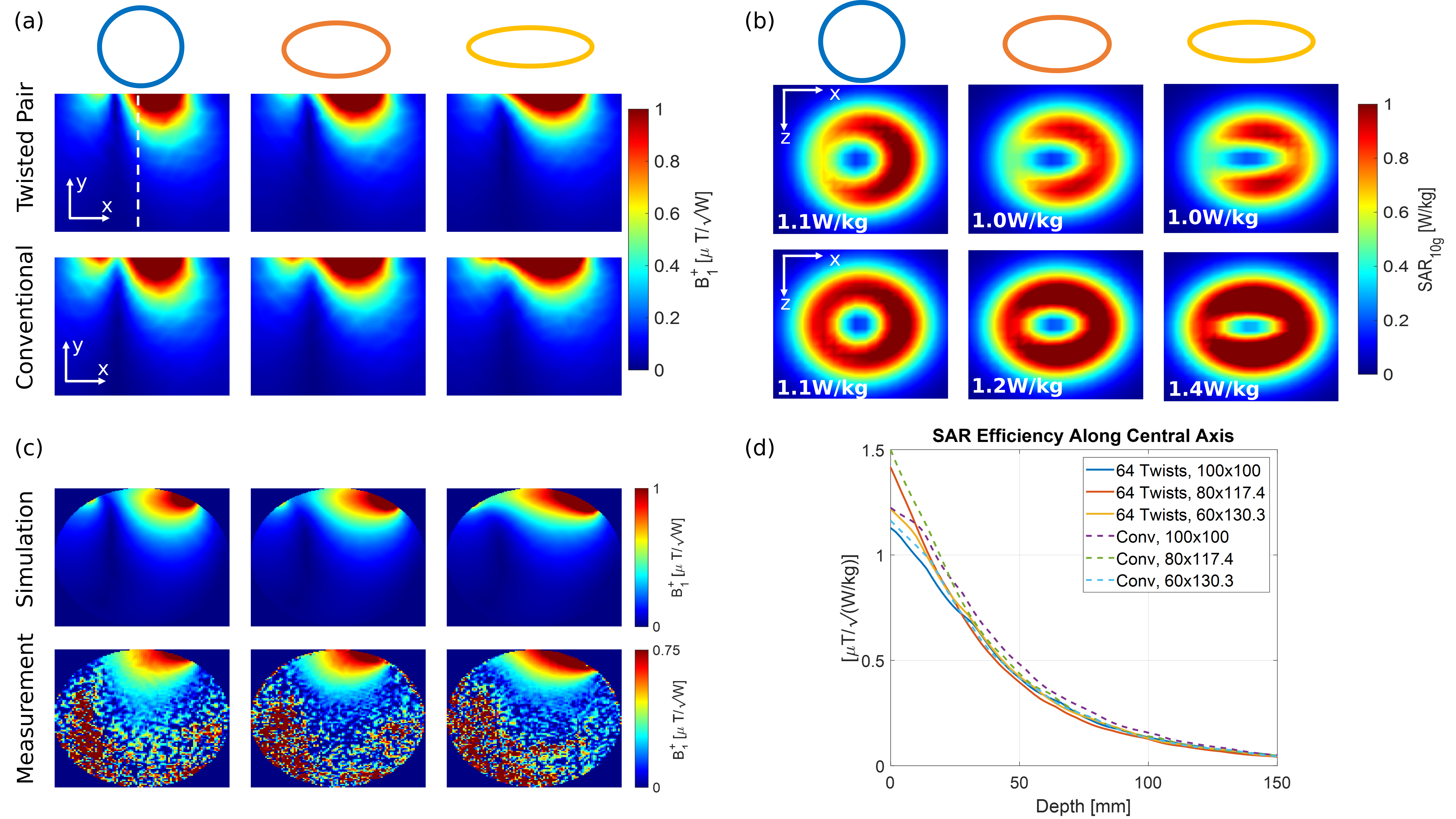}
	\caption{Comparison between the twisted pair, with 64 twists per unit length, and conventional coil, for different degrees of elongation (100$\times$100, 80$\times$117.4, 60$\times$130.3, all in mm). All results are normalised to 1~W accepted power. (a) B$_1^+$ field patterns, in $\mu \text{T}/\sqrt{\text{W}}$, on the sagittal plane. The white dashed line shows the central axis of the phantom. (b) Coronal view SAR\textsubscript{10g} maps, in W/kg, accompanied by the corresponding maxSAR\textsubscript{10g} value. (c) Simulated (top) and measured (bottom) B$_1^+$ field maps on a spherical phantom for varying degrees of elongation. The simulated results take 25\% loss of the experiment into account, while the measurements were normalised to the input power. (d) SAR efficiency: B$_1^+$ divided by the square root of peak SAR, in $\mu \text{T}/\sqrt{\text{W/kg}}$, along the central axis of the phantom.}
	\label{fig:figure4}
\end{figure}

\begin{figure}[H]
	\centering
	\captionsetup{justification=centering}
	\includegraphics[width=\textwidth]{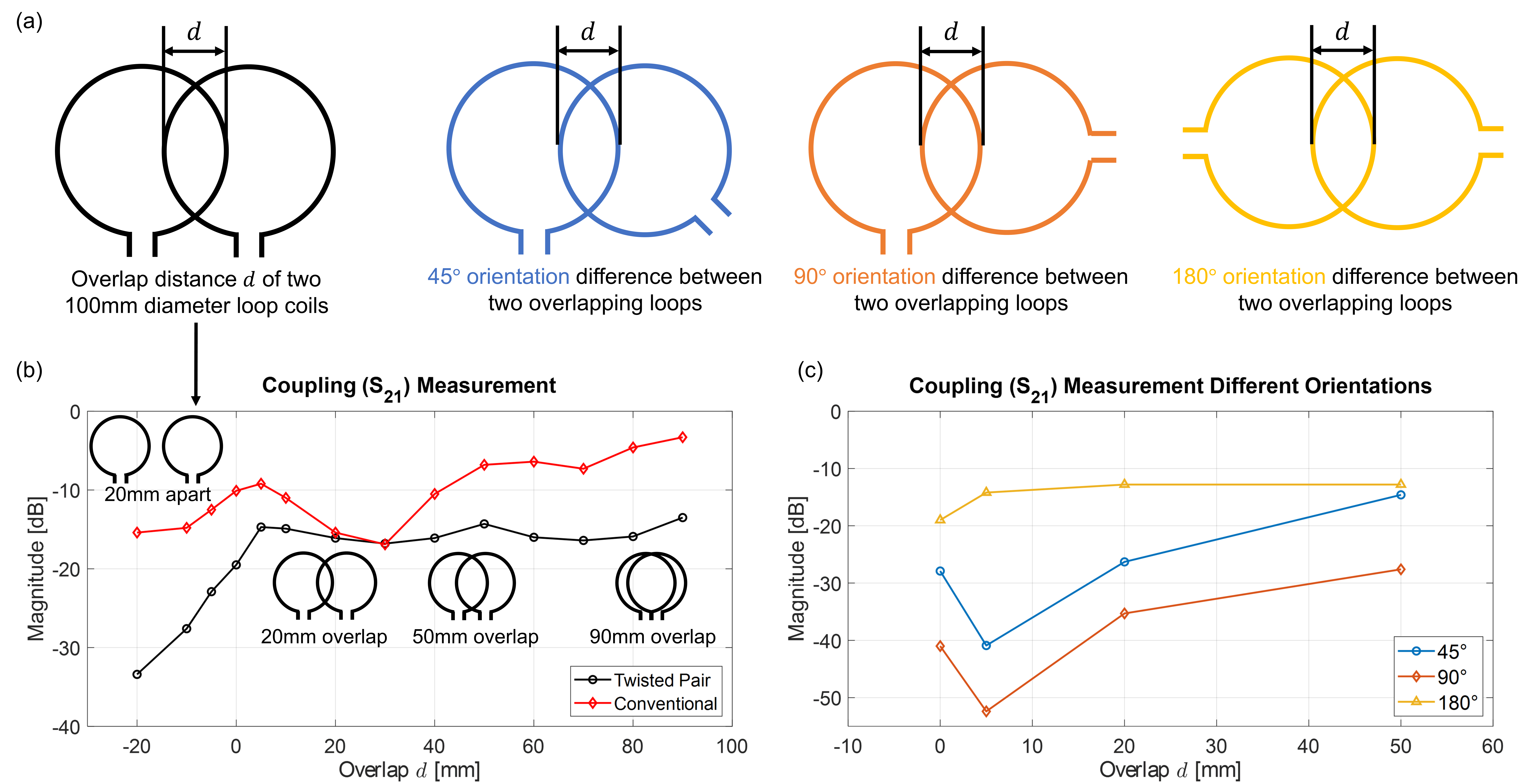}
	\caption{Measured coupling coefficients ($S_{21}$), in dB, of two coils for varying overlapping geometries. (a) Graphical representation of the different coupling orientations. (b) Coupling  results of two loops, twisted pair and conventional, next to each other for varying degrees of overlap $d$, in mm. (c) Coupling results for different orientations of the twisted pair loops, as indicated in (a).}
	\label{fig:figure7}
\end{figure}

\begin{figure}[H]
	\centering
	\captionsetup{justification=centering}
	\includegraphics[width=\columnwidth]{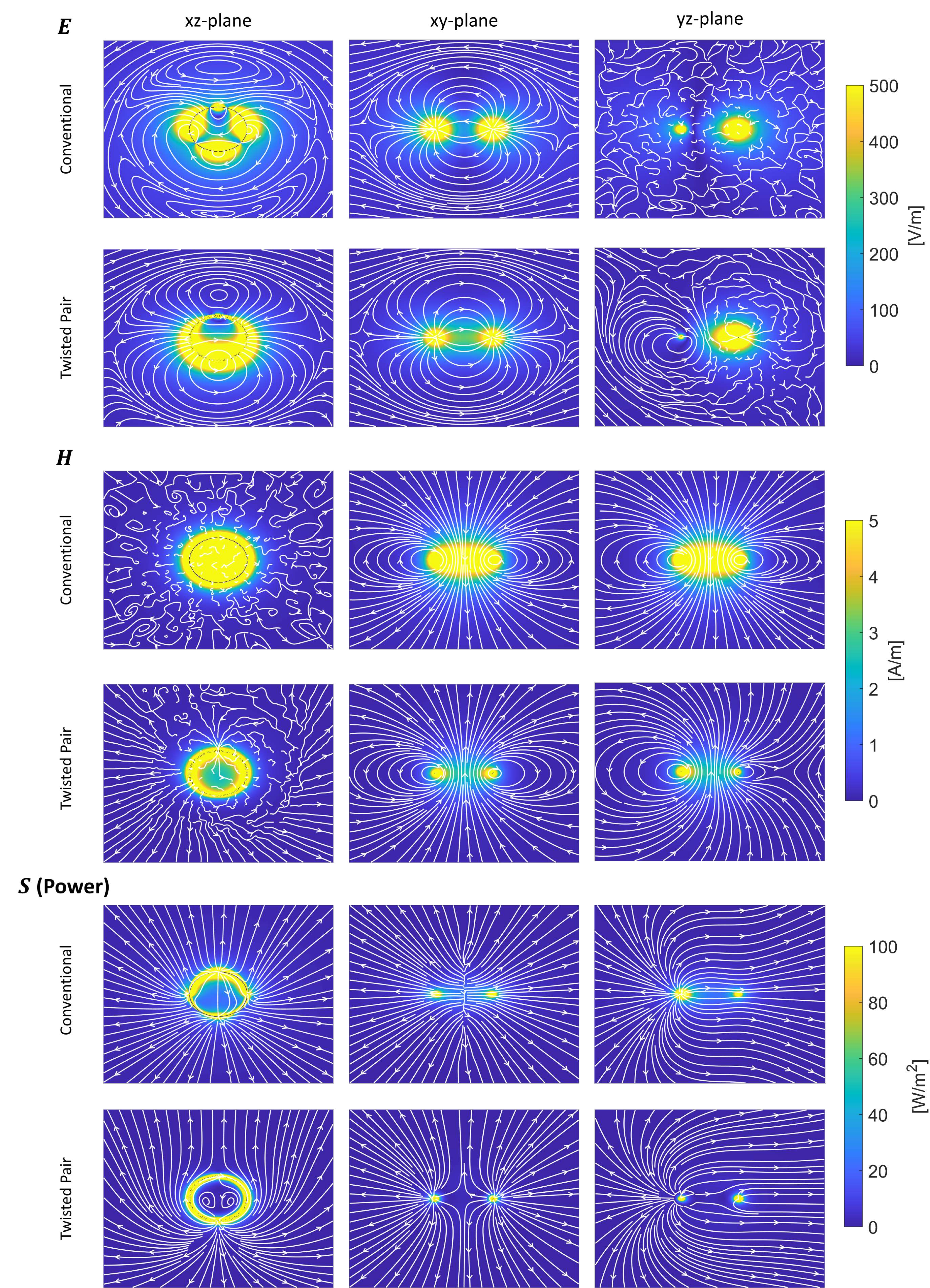}
	\caption{Streamline representation of electric ($\bm{E}$) and magnetic ($\bm{H}$) fields and power flow (Poynting vector $\bm{S}$) of individual conventional and twisted pair coils in air (no phantom present).}
	\label{fig:NoPhantom}
\end{figure}

\begin{figure}[H]
	\centering
	\captionsetup{justification=centering}
	\includegraphics[width=\columnwidth]{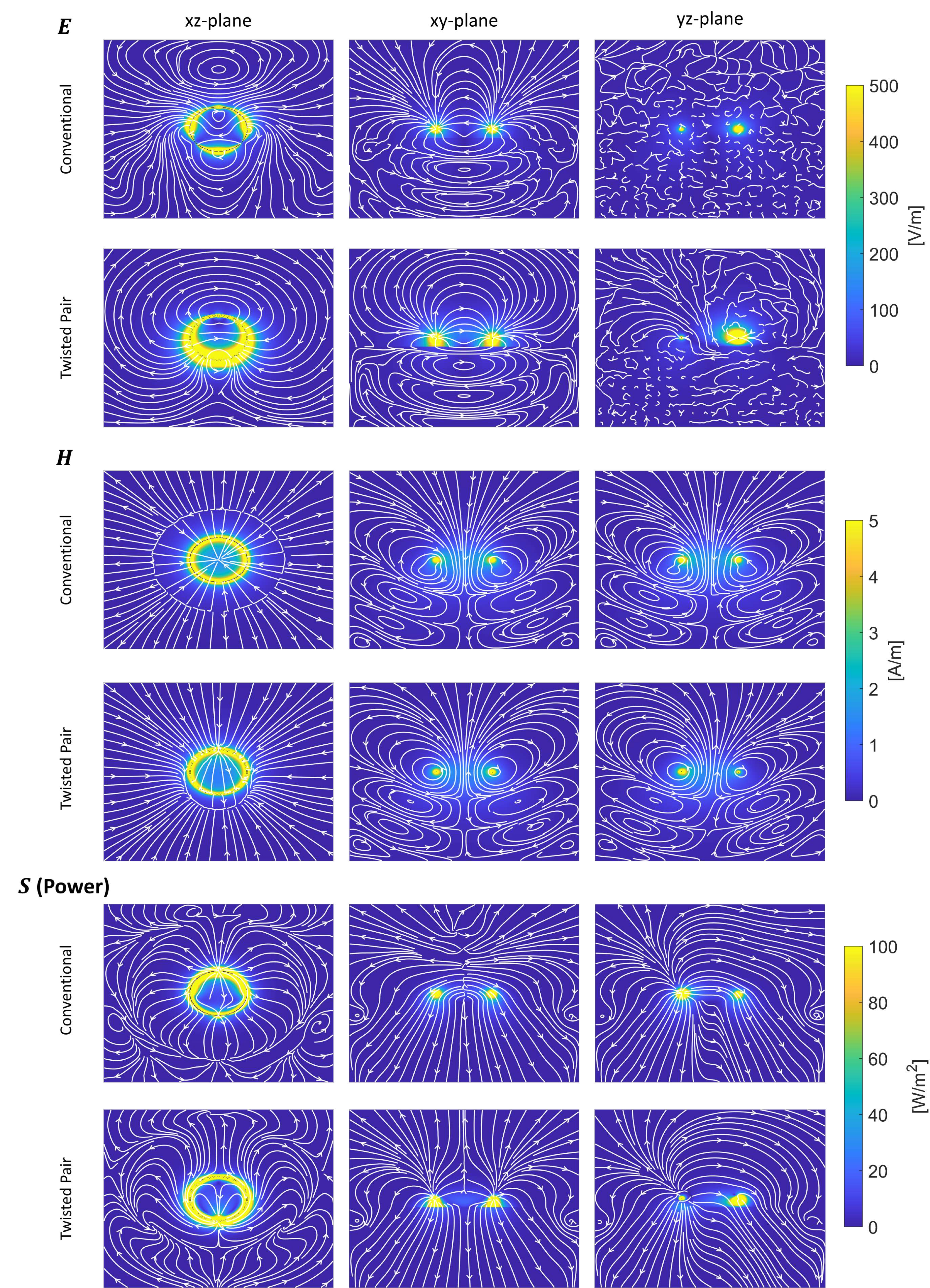}
	\caption{Streamline representation of electric ($\bm{E}$) and magnetic ($\bm{H}$) fields and power flow (Poynting vector $\bm{S}$) of individual conventional and twisted pair coils in the presence of the phantom.}
	\label{fig:WithPhantom}
\end{figure}

\begin{figure}[H]
	\centering
	\captionsetup{justification=centering}
	\includegraphics[width=\columnwidth]{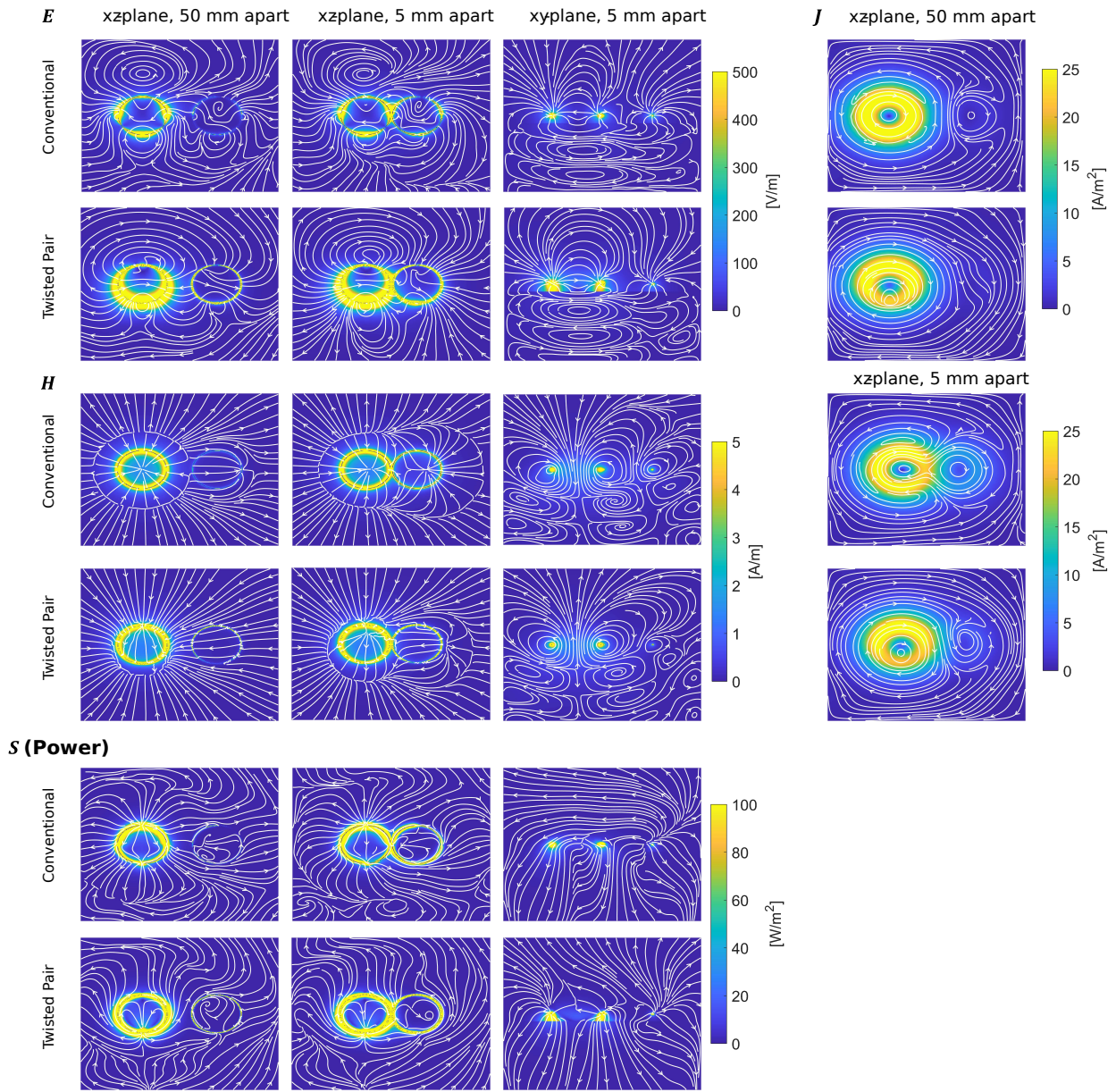}
	\caption{Streamline representation of  electric ($\bm{E}$) and magnetic ($\bm{H}$) fields, surface current density ($\bm{J}$) and power flow (Poynting vector $\bm{S}$) of two conventional and two twisted pair coils in the presence of phantom for distances of 5~cm and 5~mm from each other.}
	\label{fig:TwoWithPhantom}
\end{figure}

\begin{figure}[H]
	\centering
	\captionsetup{justification=centering}
	\includegraphics[width=\textwidth]{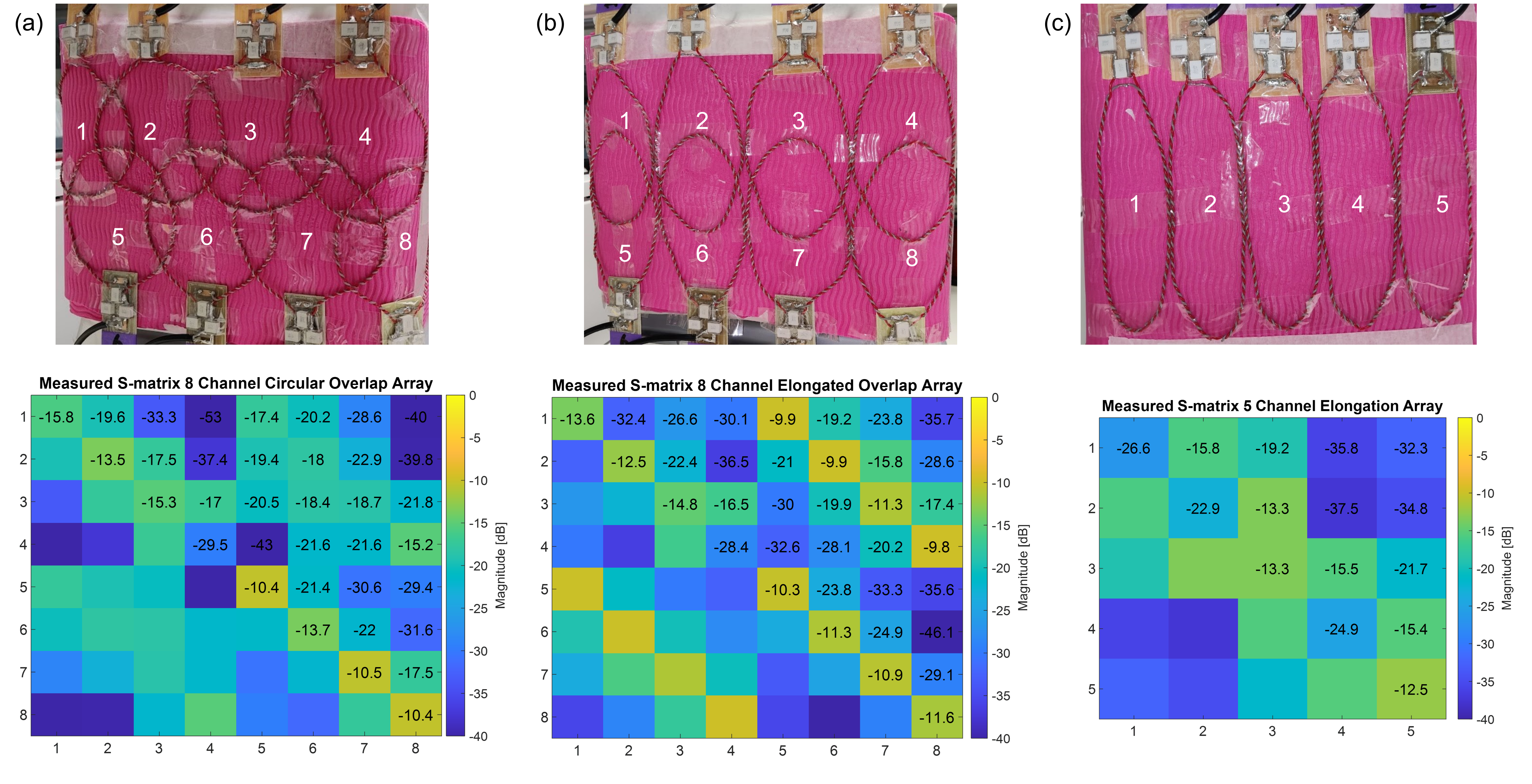}
	\caption{Measured $S$-matrices, in dB, for multiple twisted pair coil coupling layouts: (a) Eight-channel circular overlap; (b) Eight-channel elongated overlap; (c) Five channel elongated array.}
	\label{fig:figure9}
\end{figure}

\begin{figure}[H]
	\centering
	\captionsetup{justification=centering}
	\includegraphics[width=\textwidth]{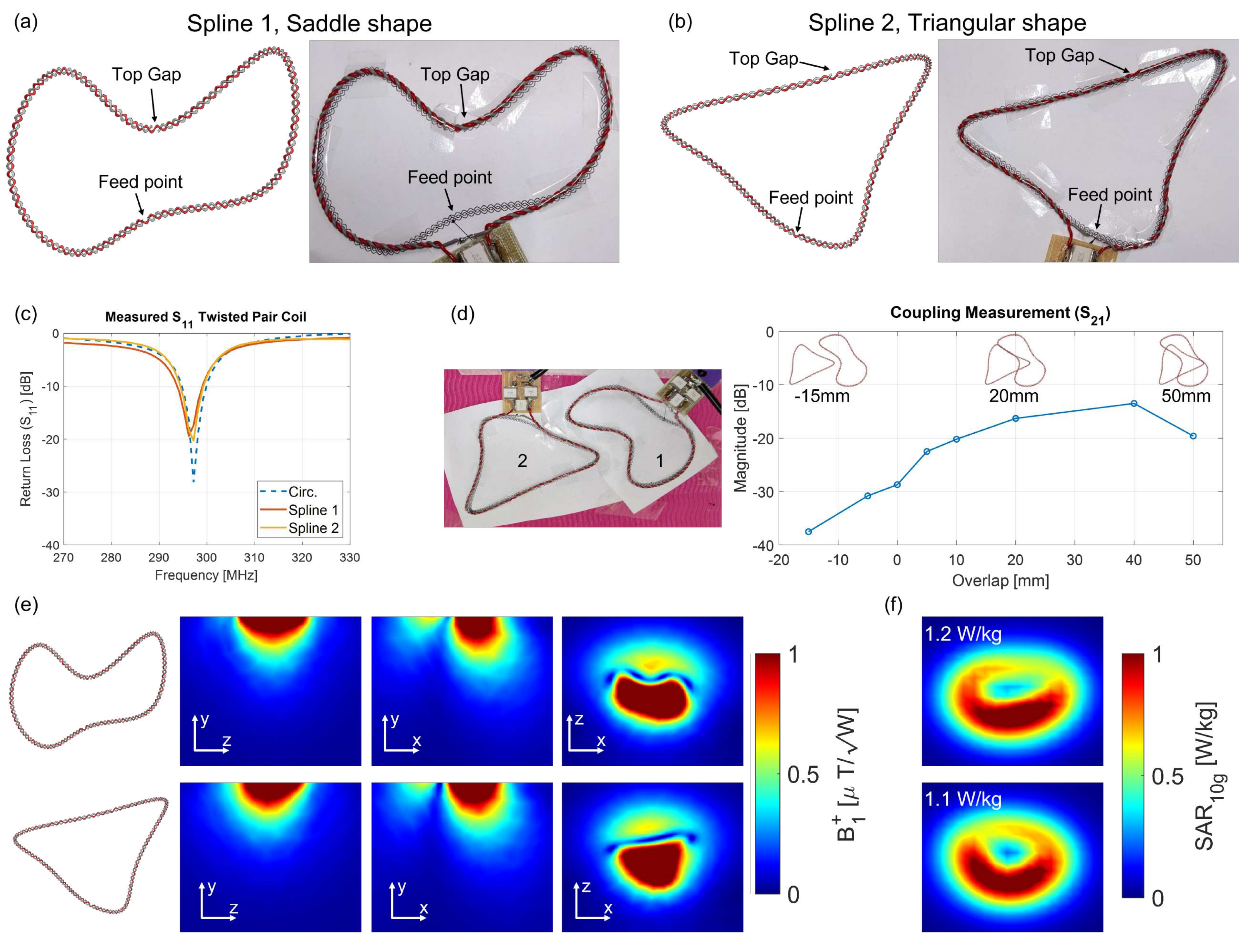}
	\caption{Different metrics for two realised spline shapes, (a) the \textit{Saddle} and (b) the \textit{Triangle}. (c) Measured return loss ($S_{11}$), in dB, for both splines and a circular twisted pair coil. (d) Coupling ($S_{21}$) between the splines for varying degrees of overlap. (e) Simulated B1+ maps, in $\mu \text{T}/\sqrt{\text{W}}$, and (f) SAR maps, in W/kg, on the coronal plane with its corresponding peak SAR value. Simulated results were normalised to 1 W accepted power}
	\label{fig:SplinesAll}
\end{figure}

Considering the design parameters depicted in \autoref{fig:figure2}, it becomes evident that a higher number of twists per unit length, such as 64, results in an inherently decoupled element with a lower maxSAR\textsubscript{10g} value in comparison to a lower number of twists per unit length, for instance, 10. As the number of twists increases, both the B$_1^+$ efficiency at a depth of 5~cm and maxSAR\textsubscript{10g} decrease, implying nearly constant SAR efficiency across different twist densities. This decline in B$_1^+$ and SAR is likely attributed to amplified EM field cancellation stemming from the increased number of twists. The incorporation of additional twists leads to a higher distributed capacitance, causing a downward shift in the resonance frequency due to the extended electrical length. This heightened capacitance simultaneously contributes to a reduction in coil impedance, as indicated in the Smith chart presented in \autoref{fig:figure2}~(a). The increased distributed capacitance also retains the electric field more, resulting in lower interelement interactions for a higher number of twists per unit length.

The cancellation of the EM field, when currents flow in opposite directions within the wires, depends on the configuration and relative orientation of these wire segments. A greater number of twists facilitates more substantial interaction between the individual wire segments’ generated fields. In contrast, a coil with fewer twists may experience less cancellation due to limited opportunities for EM field interactions, leading to a stronger field that increases both B$_1^+$ and SAR, compared to a higher twist per unit length coil. Given the similar SAR efficiency observed across coils with different twist densities, we opted to utilise the element with the highest robustness and lowest interelement coupling for subsequent simulations and experiments, specifically the configuration with 64 twists per unit length. If an element with an increased B$_1^+$ field efficiency and possibly higher Q factor is desired one could use an element with lower twists per unit length, such as 10 twists.

A valid question that might arise is why not more than 64 twists per unit length were subjected to analysis. This limitation can be attributed to two primary factors: Firstly, the wires employed in the fabrication of the coils exhibit a threshold of around 64 twists per unit length, beyond which they become susceptible to snapping. While employing alternative wires could address this concern, such wires would need to possess a smaller wire diameter to accommodate more than 64 twists per unit length. Secondly, simulations involving more than 64 twists per length for a 10 cm diameter loop coil become increasingly inaccurate. Notably, the model begins to manifest sharp edges, resulting in imprecise mesh representation.

Streamline visualisation of electric and magnetic fields and power flow, shows the coupling mechanism of the coils and their energy flow. The power is concentrated around a twisted wire structure and only a small amount is radiated away or present inside of the coil (similar results were obtained when the SCC coil was analysed by streamline approach, not shown in this manuscript). The origin of the power lines in the twisted pair coil was at the gap position (opposite side from the feed point), whereas for the conventional coil, it flows from the feed point. Very strong capacitive coupling between the coil and phantom can be noticed and resulted in a strong decrease of the transmit field efficiency when the twisted pair coil is moved away from the phantom (see \ref{appendix:APdistancePhantom}). When a conventional coil was moved away from the phantom (from a distance of 20~mm to a distance of 40~mm, for example) the transmit field efficiency reduction was much lower (8\% compared to 15\% for the twisted pair) than in the twisted pair coil because the conventional coil coupled to the phantom not only via capacitive coupling but also via radiative field. Close to the phantom, both twisted pair and conventional coils exhibited similar transmit field efficiencies. By looking at the streamline fields of two coils on a phantom setup, it can be concluded that in the case of twisted pair coils the interelement coupling occurs dominantly through the electric field – electric field lines originated at the first (active) coil and ended on a second (passive coil). This trend was clearly visible at any cut. The second (passive) twisted pair coil had little interaction with the magnetic field originating from the first coil or from the phantom, while for the conventional coil, the coupling through a phantom and directly through the magnetic field produced by the active coil was present. The dominant coupling mechanism of a twisted pair coil is through an electric field originating from the coil itself and it does not depend on a coil distance from the phantom. Unloaded and loaded Q factor measurements are almost identical which tells the twisted pair coil is not dominantly influenced by the presence of the phantom. On the other side, when the twisted pair coil is close to the phantom, there is very strong capacitive coupling between the twisted pair coil and phantom itself (very strong $\bm{E}$ field and concentrated power between the coil and phantom, \autoref{fig:WithPhantom}), which does not affect interelement coil coupling but B$_1^+$ field efficiency. Due to the capacitive nature of the twisted pair coil coupling to the phantom, there is a steeper decrease in B$_1^+$ field efficiency when the twisted pair coil is moved away from the phantom than is the case with the conventional coil. As already mentioned, twisted pair coils are less sensitive to the secondary magnetic field produced by induced currents in a phantom (so-called resistive coupling).

When looking at the streamlines of an SSC coil (not shown in this work) similar features are observed as for the twisted pair coil: they effectively guide energy within the coil structure, minimizing outward radiation; have a high capacitive coupling with the phantom; largely reduced inductive coupling component. However, this leads to the question: What advantages does the twisted pair coil offer over the SCC? While both operate on similar principles, a benefit of the twisted pair coil is the additional degrees of freedom of manufacturing (number of twists) and increased flexibility due to the absence of the dielectric material inside the coaxial cable.

All simulated and fabricated coils were tuned and matched only once and $S_{11}$ was stable while being elongated, deformed and placed in various array configurations. Supplementary material 1 shows a video of different shape deformation of the twisted pair coil element while the $S_{11}$ remains stable.  

In terms of B$_1^+$ \textit{field efficiency}, the coil has very similar performances compared to the conventional coil of the same diameter (with distributed capacitors) when placed close to the phantom. Insensitivity to the coil shape deformation and stable $S_{11}$ and SAR are the biggest advantages of the twisted pair coil compared to the conventional design. In circular form, both twisted pair and conventional coils had the same maxSAR\textsubscript{10g} of 1.1 W/kg. In elongated forms, twisted pair coils showed decreased SAR from 1.1 W/kg to 1 W/kg, while the SAR of the conventional coils increased by 30\% at maximum elongation. Deformation of twisted pair coil to various spline shapes produced a minimal increase of SAR (1.1 W/kg and 1.2 W/kg, for the two example shapes). 

When examined in terms of \textit{interelement decoupling properties}, the twisted pair coil showed a truly self-decoupled nature – it stayed decoupled (better than -10 dB) in all examined array configurations without the need for additional decoupling techniques. It is shown that even in extreme coil shape deformations (spline coil shapes shown in\autoref{fig:SplinesAll}), the $S_{11}$ and SAR of individual coils are stable, and remain well decoupled when placed at various distances from each other. Such superior decoupling properties make twisted pair transmission line coil a suitable element for building densely populated tight-fit receive arrays. This leads to the idea to optimise the shape of the coil when placed in receive array configuration in such a way as to allow high accelerations and maximally improved g-factor.  

An additional benefit of the twisted pair over the SCC coil is the absence of a thicker, inflexible dielectric material. This particular design feature makes the twisted pair coil well-suited for accommodating extreme shape deformations. Its robustness in handling geometric changes positions the twisted pair coil as an excellent element for optimised receive arrays for highly accelerated MRI.

There are also commercially available flexible arrays such as AIR coil element~\cite{Vasanawala2017, McGee2018, Collick2020}. The AIR coil element has been proposed as a flexible and highly decoupled element mainly in receive arrays for operation at 3T. A variety of coil positioning in array configurations and coil shapes different than circular, were not reported. Additionally, noise-controlling preamplifiers have been placed at each loop coil, which could also help improve the interelement coupling of AIR coils~\cite{Vasanawala2017}, while twisted pair transmission line coils do not need any additional decoupling technique to be used.

\section{Conclusion}\label{sec:conclusion}
This work introduced and investigated the twisted pair transmission line coil as a novel, highly flexible and self-decoupled transceive element for 7T MRI. The twisted pair transmission line coil showed comparable transmit field efficiency (B$_1^+$) and SAR performance to a conventional copper loop coil and exceptional robustness to various shape deformations. The dominant coupling mechanism is through electric fields originating from a coil itself while the coils are insensitive to the magnetic fields (both direct field originating from a coil and secondary magnetic field originating from a phantom). In this paper, a streamline approach to visualise the electric and magnetic fields and power flow has been proposed for a deeper understanding of coil functioning and the same approach can be applied in studying any other coil designs.

Overall, the findings of this study support the potential of the twisted pair transmission line coil as a highly flexible and efficient solution for 7T MRI, especially in tight-fit arrays.

\section*{Acknowledgements}
The authors would like to thank Matthijs van Osch (LUMC, Leiden, Netherlands) for his help with the measurements.

 \bibliographystyle{elsarticle-num} 
 \biboptions{sort&compress}
 \bibliography{references}

\newpage

\newpage
\appendix

\section{Working of the twisted pair} \label{appendix:APworkings}

\begin{figure}[H]
	\centering
	\captionsetup{justification=centering}
	\includegraphics[width=\textwidth]{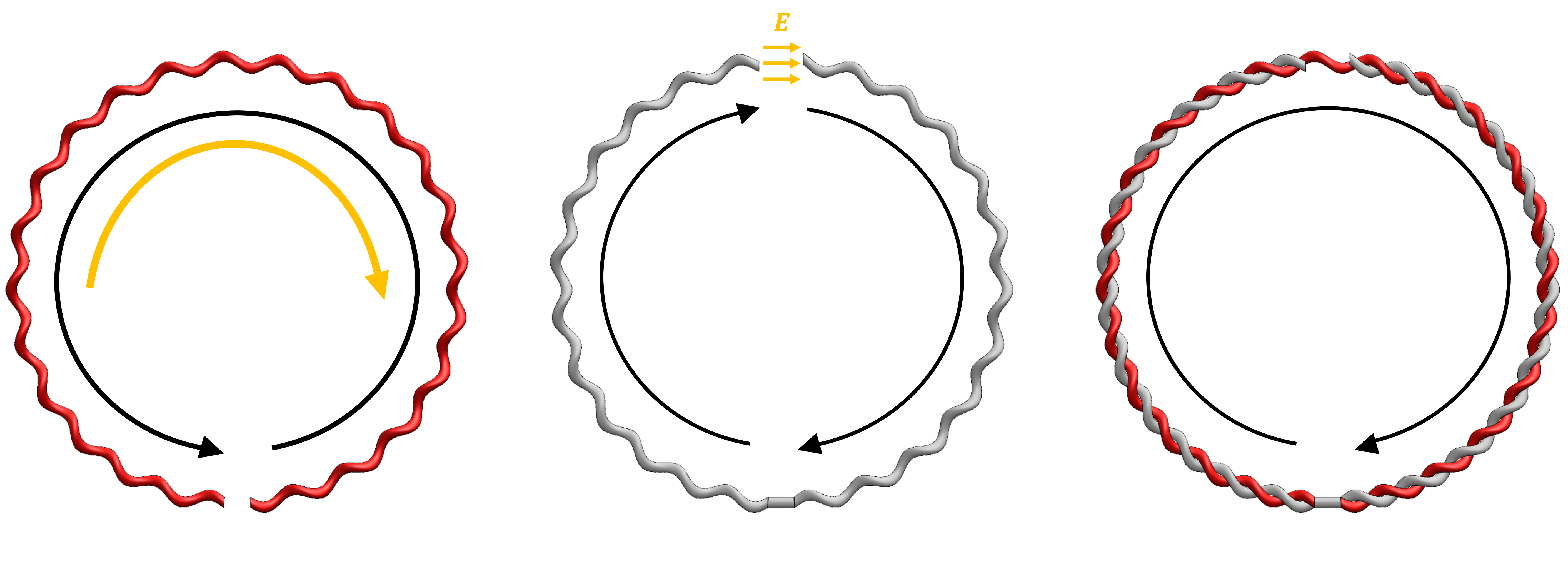}
	\caption{Illustration of the current flow on the twisted pair coil. A current is excited on the red \textit{signal wire} counterclockwise (CCW), resulting in a clockwise (CW) induced current on the grey \textit{shield wire}. At the gap of the grey \textit{shield wire} a high potential difference creates a strong electric field, which induces a current on the red \textit{signal wire} flowing in the opposite direction as the original current. As a result of this induced current, there is a partial cancellation of the current in the red \textit{signal wire}. Consequently, the grey \textit{shield wire} becomes the predominant carrier of current, as illustrated in the combined wires figure on the right (current flowing CW).}
	\label{fig:AP1pre}
\end{figure}

\section{Simulated robustness of the twisted pair} \label{appendix:AProbustness}

\begin{figure}[H]
	\centering
	\captionsetup{justification=centering}
	\includegraphics[width=\textwidth]{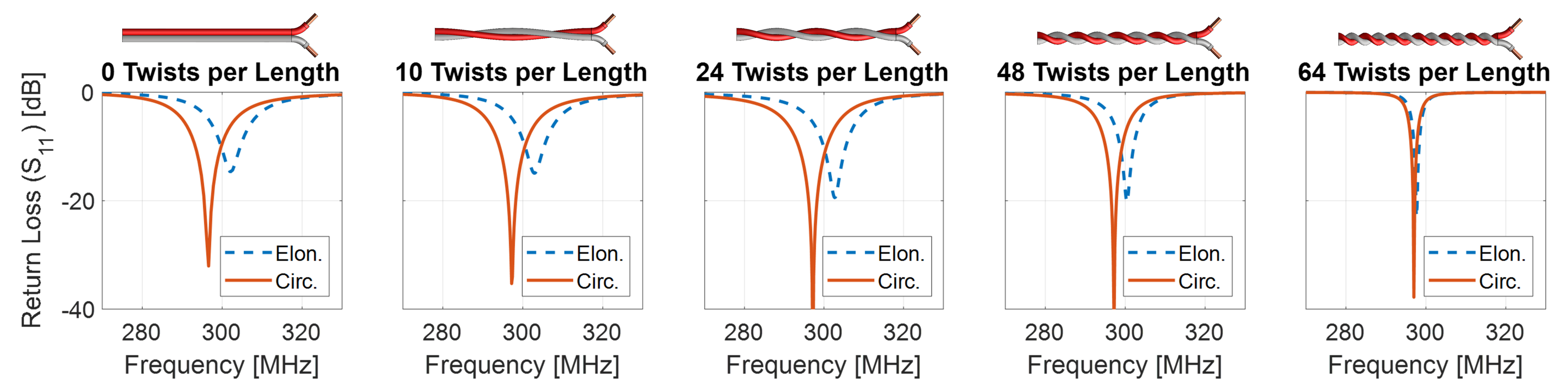}
	\caption{Simulated return loss ($S_{11}$) across varying numbers of twists per unit length. The circular coil (10 cm diameter) underwent tuning and matching, and the same circuitry was applied to the elongated coil (dimensions: width = 6 cm, height = 13 cm) to assess coil robustness.}
	\label{fig:AP1}
\end{figure}

\section{Measured robustness of the twisted pair} \label{appendix:APmeasured}

\begin{figure}[H]
	\centering
	\captionsetup{justification=centering}
	\includegraphics[width=\textwidth]{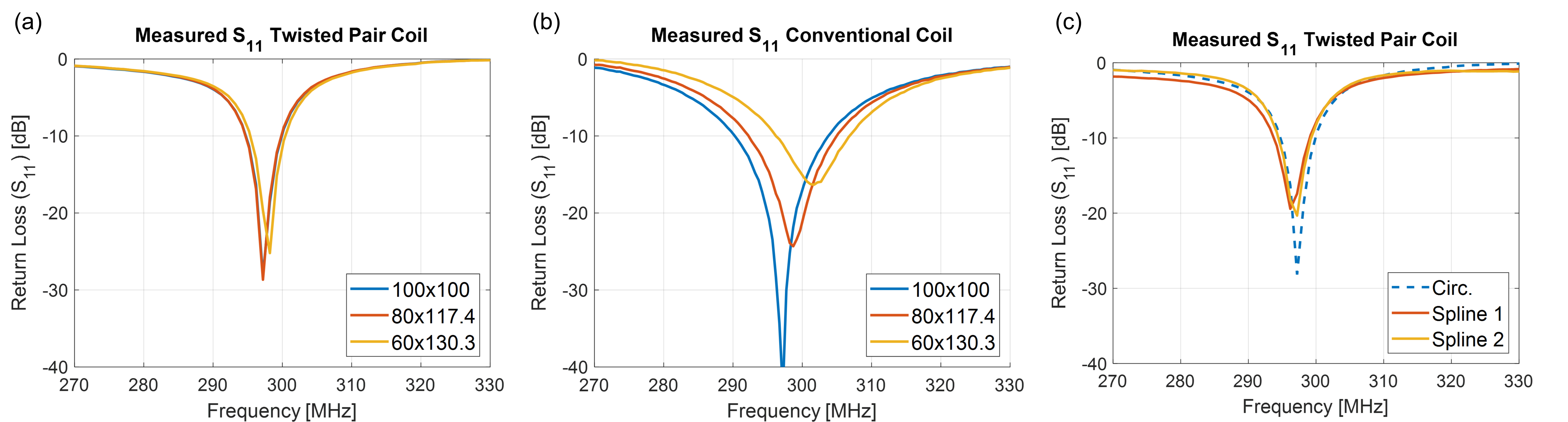}
	\caption{Return loss ($S_{11}$) measurements. All coils were initially tuned and matched in the circular geometry and subsequently subjected to deformation using identical circuitry. (a) \& (b) display $S_{11}$ for varying degrees of elongation for the twisted pair (64 twists per unit length) and the conventional coil, respectively. Panel (c) illustrates the $S_{11}$ of two spline shapes, namely \textit{saddle} and \textit{triangular}, with the circular geometry serving as the reference.}
	\label{fig:AP2}
\end{figure}

\section{Influence coil distance to the phantom} \label{appendix:APdistancePhantom}

\begin{figure}[H]
	\centering
	\captionsetup{justification=centering}
	\includegraphics[width=\textwidth]{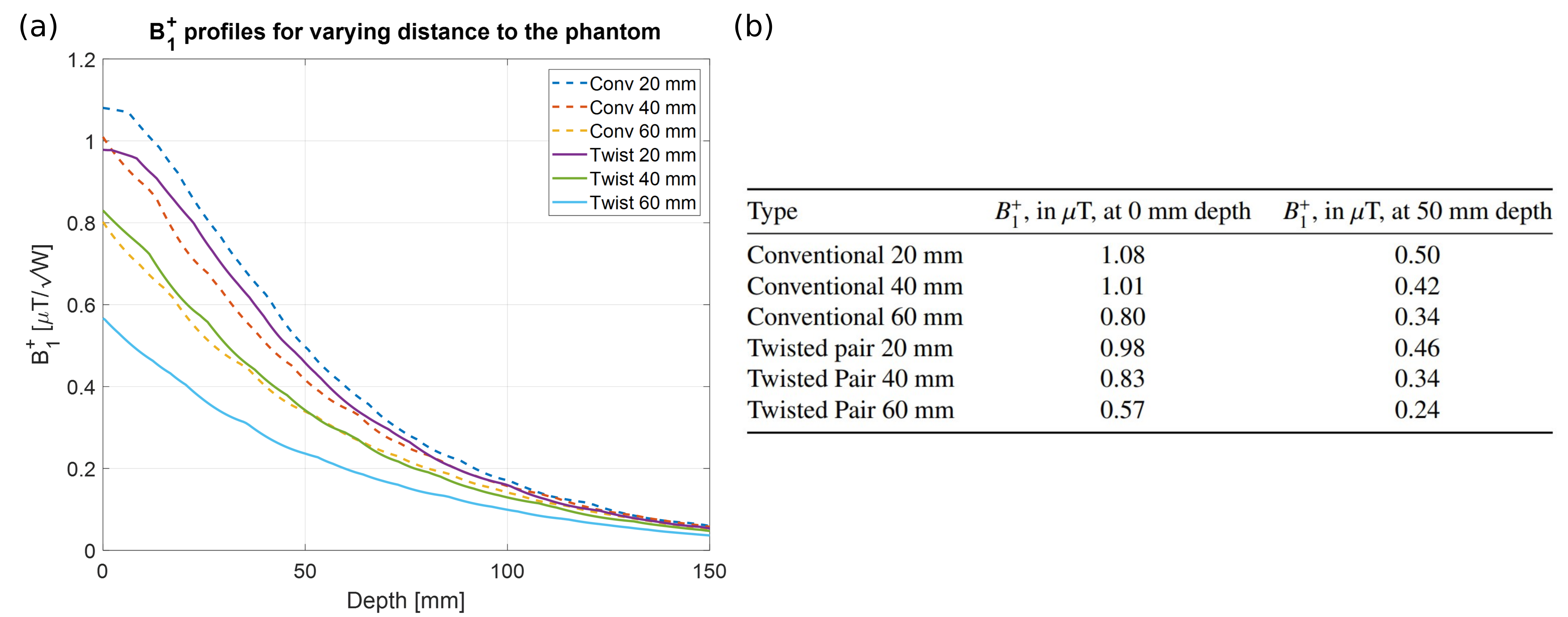}
	\caption{Influence on the simulated B$_1^+$ for the distance of the coil to the phantom. (a) The B$_1^+$ profiles are plotted along the central axis of the phantom for the conventional, dashed line, and the twisted pair, full line. Both coils were placed at 20, 40 and 60 mm. (b) Simulated magnitude values at the surface of the phantom (0 mm) and 50 mm depth.}
	\label{fig:AP3}
\end{figure}

\end{document}